# Impact of May 2024 ICME Event on Martian UV Dayglow Emissions using MAVEN/IUVS and EMM/EMUS Observations

Aadarsh Raj Sharma[1], Vaghamshi Dhruv Dhaneshbhai [1], Lot Ram[2], Sumanta Sarkhel[1,3*]

*Sumanta Sarkhel, Department of Physics, Indian Institute of Technology Roorkee, Roorkee - 247667, Uttarakhand, India (sarkhel@ph.iitr.ac.in)

[1]Department of Physics, Indian Institute of Technology Roorkee, Roorkee, Uttarakhand, India

[2]JSPS International Research Fellow, Graduate School of Frontier Sciences, The University of Tokyo, Kashiwa, Japan

[3]Centre for Space Science and Technology, Indian Institute of Technology Roorkee, Roorkee, Uttarakhand, India

**Abstract**

We present first near-simultaneous observations of large-scale spatial response of the Martian H Lyman-α and atomic oxygen dayglow emissions (O($^3$S) 130.4 nm and O($^5$S) 135.6 nm) during May 2024 interplanetary coronal mass ejection (ICME). Using coordinated near-simultaneous observations from MAVEN/Imaging Ultraviolet Spectrograph (IUVS) and the EMM/Emirates Mars Ultraviolet Spectrometer (EMUS), we have identified pronounced enhancements in all three emissions relative to quiet-time conditions. Interestingly, coordinated IUVS limb and EMUS disk observations reveal that ICME-driven response extends vertically through the thermosphere and horizontally across broad ranges of latitude, longitude, local-time, and solar zenith angle. Concurrent enhancement of hydrogen and oxygen emissions indicates the simultaneous contribution of proton and electron-driven excitation mechanisms during ICME. These findings reveal a novel planetary-scale response of the Martian dayglow. By highlighting the critical role of energetic-particle precipitation in amplifying this global response, our work carries crucial implications for understanding Martian thermospheric response during intense solar transients.

**Key Points:**

1. First near-simultaneous MAVEN/IUVS and EMM/EMUS observations reveal large-scale ICME driven enhancement in Martian UV dayglow emissions.
2. Simultaneous H Lyman-α and atomic oxygen emissions enhancements indicate concurrent proton and electron precipitation during the ICME.
3. Increased contribution of electron-impact excitation due to energetic particle precipitation caused the enhancements in dayglow emissions.

## Plain Language Summary

The H Lyman-α, O($^3$S) 130.4 nm, and O($^5$S) 135.6 nm emissions are among the most important ultraviolet dayglow emissions in the Martian upper atmosphere. H Lyman-α is primarily produced through resonance scattering of solar photons by coronal hydrogen, while the O($^3$S) 130.4 nm and O($^5$S) 135.6 nm emissions are produced through resonance scattering, photodissociation, and electron impact excitation processes. Space weather events such as interplanetary coronal mass ejections (ICMEs) can significantly perturb the Martian ionosphere and thermosphere through enhanced proton and electron precipitations. This motivates us to investigate the response of these ultraviolet emissions during extreme ICME conditions. In this study, we report for the first time, the large-scale spatial response of H Lyman-α, O($^3$S) 130.4 nm, and O($^5$S) 135.6 nm emissions during the May 2024 ICME event using observations from MAVEN/IUVS and EMM/EMUS. Our results show significant enhancements in all three emissions during the ICME compared to quiet-time conditions. The concurrent enhancement of hydrogen and oxygen emissions indicates simultaneous operation of multiple particle-driven excitation mechanisms during ICME. Despite the significant increase in brightness, the characteristic peak emission altitudes and large-scale spatial distributions remain largely unchanged, suggesting that the ICME primarily intensifies existing emission spatial structures.

## 1. Introduction

The Martian upper atmosphere has been probed at ultraviolet (UV) wavelengths since the Mariner 6, 7, and 9 missions, which initially detected H Lyman-α (121.6 nm), the O($^3$S) 130.4 nm and O($^5$S) 135.6 nm UV dayglow emissions (Barth et. al., 1971). The H Lyman-α is produced predominantly by resonance scattering of solar photons by coronal hydrogen, with >90% of quiet-time dayside brightness arising from this channel (Chaffin et al., 2015). Similarly, O($^3$S) 130.4 nm dayglow emission is majorly formed by resonant scattering (~80-85%; R1) with a contribution of ~15-20% production via the electron impact on O and $CO_2$ species (R2 & R3; Leblanc et. al., 2006; Ritter et. al., 2019).

$$h\nu_{130.4} + \mathrm{O} \rightarrow \mathrm{O}(^3S) \qquad \text{(R1)}$$

$$\mathrm{e}^- + \mathrm{O} \rightarrow \mathrm{O}(^3S) + \mathrm{e}^- \qquad \text{(R2)}$$

$$\mathrm{e}^- + \mathrm{CO_2} \rightarrow \mathrm{CO} + \mathrm{O}(^3S) + \mathrm{e}^- \qquad \text{(R3)}$$

Whereas the O($^5$S) 135.6 nm dayglow emission is dominantly produced by the electron impact excitation of O (95%; R4), with a minor contribution from dissociative excitation of $CO_2$ (R5; Leblanc et. al., 2006; Ritter et. al., 2019).

$$\mathrm{e}^- + \mathrm{O} \rightarrow \mathrm{O}(^5S) + \mathrm{e}^- \qquad \text{(R4)}$$

$$\mathrm{e}^- + \mathrm{CO_2} \rightarrow \mathrm{CO} + \mathrm{O}(^5S) + \mathrm{e}^- \qquad \text{(R5)}$$

Therefore, the relative behavior of the O($^3$S) 130.4 nm and O($^5$S) 135.6 nm emissions provides a sensitive diagnostic of photon-driven versus particle-driven excitation processes in the Martian upper atmosphere. The emission peak of O($^3$S) 130.4 nm and O($^5$S) 135.6 nm emissions usually exist at ~130-150 km altitude (Ritter et. al., 2019). Previous studies have characterized the morphology, altitude distribution, seasonal variability, and solar cycle dependence of Martian UV dayglow emissions under nominal conditions (Leblanc et. al., 2006; Chaffin et. al., 2015; Ritter et. al., 2019).

In addition to long-term variability, space weather events (e.g., ICMEs, CIRs, and Solar Flares) can cause significant modulations in Martian ionospheric peak, depletion in the dayside and nightside plasma density (Jakosky, Grebowsky, et al., 2015; Ram et al., 2023; Thampi et al., 2021; Ram, Rout, et. al., 2024; Parrott et. al., 2026; Pranjali et. al., 2026). Enhanced UV emissions have been previously reported during space weather events (Deighan et. al., 2018; Ritter et. al., 2018; Ram, Sharma et. al., 2024; Sharma et. al., 2025), indicating that space weather can substantially modify atmospheric excitation processes. Among the UV emissions, H Lyman-α exhibits an additional space-weather response through proton aurora, which forms when solar wind protons undergo charge exchange with exospheric hydrogen to produce energetic neutral atoms (ENAs) that subsequently precipitate into the atmosphere and excite hydrogen emissions (R6; Deighan et al., 2018; Chaffin et al., 2022).

$$\mathrm{H}_{sw}^{+} + \mathrm{H}_{exo} \rightarrow \mathrm{H}_{ENA} + \mathrm{H}^{+},$$

$$\mathrm{H}_{ENA} + \mathrm{CO}_2 \rightarrow \mathrm{H}^{*} + \mathrm{CO}_2,$$

$$\mathrm{H}^{*} \rightarrow \mathrm{H} + h\nu_{121.6\,nm} \qquad \text{(R6)}$$

Numerous studies have shown proton aurora and its dependence on solar wind conditions, magnetic topology, and seasonal variability (Hughes et al., 2019, 2023, 2025, 2026; Chaffin et al., 2021, 2022). However, most studies have focused on individual emissions, auroral signatures, or single-instrument observations, limiting characterization of the global and altitude-dependent UV dayglow response during ICME. Consequently, changes in dayglow brightness, vertical extent, spatial structure, and the relative roles of enhanced solar illumination, proton precipitation, and electron precipitation remain poorly constrained.

In this study, we use near-simultaneous MAVEN/IUVS limb and EMM/EMUS disk observations during the extreme May 2024 ICME to investigate the response of H Lyman-α, O($^3$S) 130.4 nm, and O($^5$S) 135.6 nm emissions. For the first time, the combined dataset enables simultaneous characterization of the vertical structure and horizontal extent of the ICME response, providing new constraints on the processes driving Martian UV dayglow enhancement.

## 2. Data and Methodology

The data used in this study are obtained from remote sensing and in situ instruments onboard the Mars Atmosphere and Volatile EvolutioN (MAVEN; Jakosky et al., 2015) and Emirates Mars Mission (EMM; Almatroushi et al., 2021; Amiri et al., 2022) spacecrafts. MAVEN operates in an elliptical orbit with a period of ~3.5 hr, apoapsis of ~6200 km, periapsis altitude of ~180 km, and inclination of ~75°. Whereas EMM orbits Mars with a period of ~55 hr in a ~20,000 × 43,000 km orbit at an inclination of ~25°, enabling near large-scale coverage of the Martian disk.

Observations of H Lyman-α, O($^3$S) 130.4 nm, and O($^5$S) 135.6 nm dayglow emissions are obtained using the Imaging Ultraviolet Spectrograph (IUVS)(McClintock et al., 2015) onboard MAVEN and the Emirates Mars Ultraviolet Spectrometer (EMUS) (Holsclaw et al., 2021) onboard EMM. We use IUVS periapsis limb scans, which provide altitude resolved emission profiles across multiple tangent points with a vertical resolution of ~5 km. In addition, EMUS disk observations obtained in U-OSp mode is used to complement the IUVS limb observations by providing large-scale spatial coverage of the UV dayglow during the ICME event.

The upstream solar wind parameters, including proton density, velocity, and dynamic pressure, as well as the interplanetary magnetic field magnitude (IMF; |**B**|), are obtained from MAVEN instruments including the Solar Wind Ion Analyzer (SWIA), Magnetometer (MAG), and Solar Wind Electron Analyzer (SWEA) (Halekas et al., 2015; Connerney et al., 2015; Mitchell et al., 2016). Solar energetic particle (SEP) fluxes are obtained from the SEP instrument (Larson et al., 2015). These measurements are used to identify and characterize the ICME conditions near Mars.

The ICME event is identified using the Space Weather Database of Notifications, Knowledge, Information (DONKI) archive. The IUVS and EMUS observations obtained during the ICME interval are considered as "event-time" observations and compared with corresponding "quiet-time" observations to assess the ICME driven variability. The IUVS quiet-time orbits are selected such that they share similar SZAs and spatial coverage as during the event-time observations (Figure S1 in Supporting Information). Similarly, for EMUS observations, those disk scans are utilized which share similar SZAs and spatial coverage during the quiet and event-times (Figure S2 in Supporting Information). This approach minimizes geometrical biases and isolates the ICME driven enhancement in the UV dayglow emissions.

## 3. Results

Figure 1 shows the ENLIL Cone model simulation of the 17–18 May 2024 ICME arrival at Mars (left) and the temporal evolution of upstream solar wind and energetic particle parameters measured by MAVEN between 20 April and 20 May 2024 (right). The right panels show variations in (a) solar wind density, (b) velocity, (c) solar wind dynamic pressure (SWDP), (d) interplanetary magnetic field magnitude (|**B**|), (e) solar wind electron flux, (f) SEP electron energy flux, and (g) SEP ion energy flux. The upstream solar wind intervals were identified following the algorithm provided in Halekas et al. (2017) and Ram et al. (2023). All parameters exhibited a pronounced enhancement that began at ~13:40 UT on 17 May and peaked on 18

May 2024, indicating the arrival of an intense ICME at Mars. During event, the solar wind density, velocity, SWDP, and |**B**| increased to peak magnitude of ~28 cm$^{-3}$, ~620 km s$^{-1}$, ~15 nPa, and ~45 nT, respectively. Simultaneously, the solar wind electron flux and SEP electron and ion energy fluxes magnitude increased to ~2×10$^{8}$, ~1.5×10$^{4}$, and ~5×10$^{4}$ eV cm$^{-2}$ s$^{-1}$ sr$^{-1}$, respectively. The blue and red shaded regions denote the quiet-time (orbits 21212–21218) and event-time (orbits 21219–21227) MAVEN/IUVS observations, while the dashed blue and red lines indicate the corresponding EMUS observations. The grey shaded region marks intervals with no available IUVS periapsis limb observations.

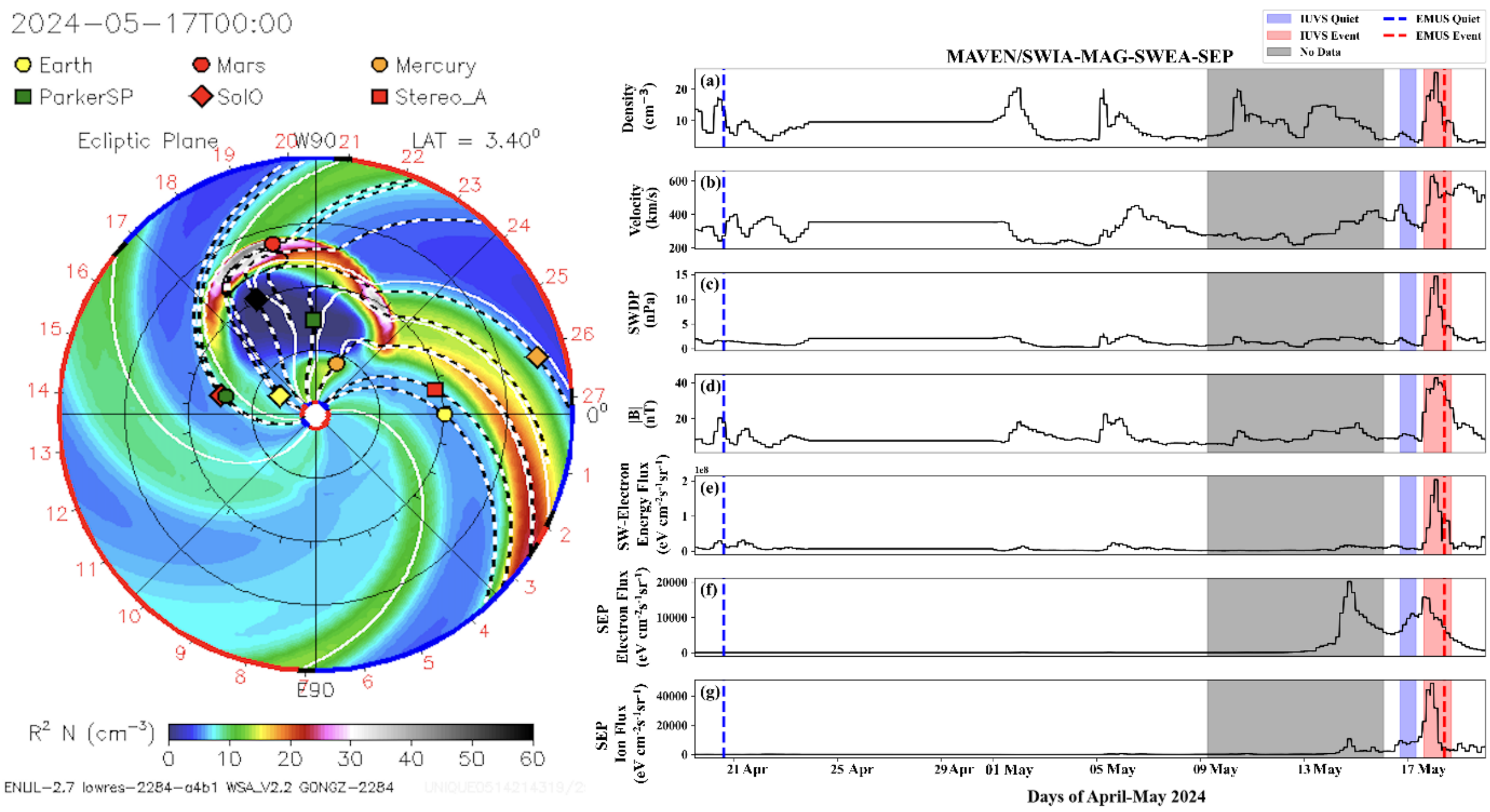

**Figure 1:** (Left) ENLIL-CONE model simulation of 17-18 May 2024 ICME event. (Right) Solar wind variations near Mars (20 April – 20 May 2024): (a) density, (b) velocity, (c) dynamic pressure (SWDP), (d) IMF (|**B**|), (e) electron density, (f) SEP electron energy flux, and (g) SEP ion density flux. The red and blue dashed line shows EMM/EMUS observation, while red and blue shaded regions show MAVEN/IUVS observations during event and quite times, respectively. Grey shaded region is referring the non-availability of IUVS limb observation.

Figure 2 shows altitude profiles of H Lyman-α, O($^{3}$S) 130.4 nm, and O($^{5}$S) 135.6 nm dayglow emissions, with brightness (kR) as a function of tangent-altitude from 70 to 200 km. Blue curves denote the mean quiet-time profiles from orbits 21212–21218, while red curves represent the mean event-time profiles from orbits 21219–21227, with shaded regions indicating 1σ variability (Figure 1). Green curves represent the orbit 21225 (18 May 2024, ~09:29 UT) during which the strongest ICME impact was observed. Only representative limb scans are shown here, while the complete set of scans are provided in Figures S3–S5 of the Supplementary Information. Additionally, the IUVS observations were largely sampled in the northern hemisphere (Figure S1 in Supporting Information).

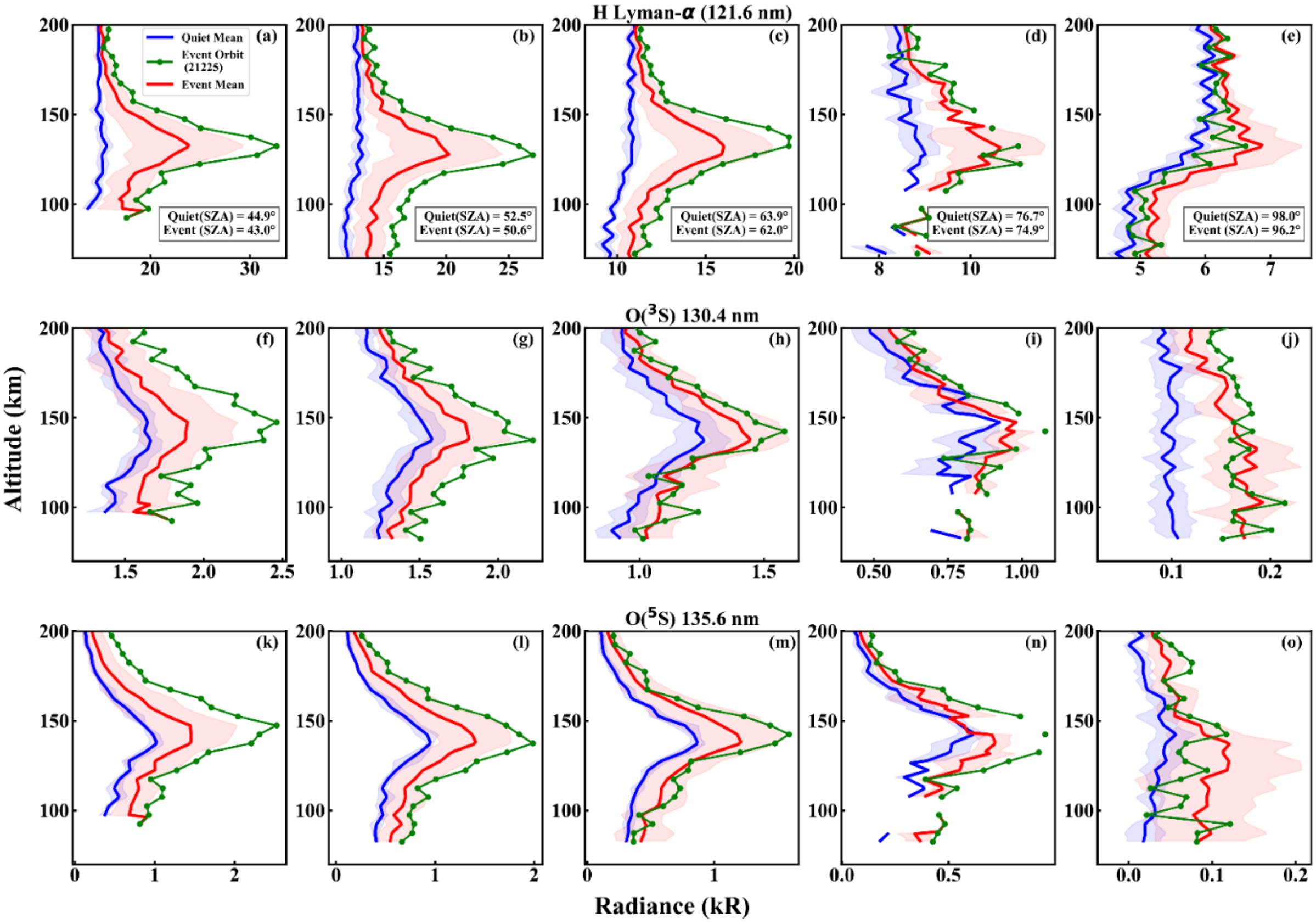

**Figure 2:** IUVS limb profiles of H Lyman-α, O(³S) 130.4 nm, and O(⁵S) 135.6 nm dayglow emissions. Red, blue, and green curves denote event-time, quiet-time, and MAVEN orbit 21225, respectively; shaded regions indicate 1σ variability.

The top-row shows H Lyman-α limb profiles for increasing SZA from left to right. At low SZA (43°-62°; Figure 2a-2c), the event-time profiles exhibit substantial enhancement relative to quite-time. Peak intensities rose from ~16 kR during quiet-time to ~24 kR during the event, reaching a maximum magnitude of ~32 kR during orbit 21225. This enhancement is strongest within the altitude range of ~110-150 km, consistent with the characteristic altitude range of proton aurora (Deighan et al., 2018; Hughes et al., 2026). With increasing SZA (Figure 2d-2e), the overall brightness decreases; nevertheless, the brightness magnitude during event-time remained consistently higher than quiet-time, suggesting that ICME-driven enhancement persisted toward the terminator.

The middle-row shows the O(³S) 130.4 nm limb emission profiles. Like H Lyman-α, these profiles exhibit a consistent enhancement during the event-time across the entire observed altitude range (70–200 km) and for all SZA ranges. The most pronounced increase occurred near the emission peak altitude (~130–150 km), where intensities rose ~1.0–1.8 kR to ~1.4–2.0 kR, ultimately reaching ~2.46 kR during orbit 21225. This enhancement persisted from the low SZA regions all the way toward the terminator. Similarly, the O(⁵S) 135.6 nm emission profiles (bottom-row) show peak intensities that increased from ~0.5–1.5 kR during quiet-time to ~0.8–1.8 kR during the event, attaining ~2.52 kR during orbit 21225. Additionally, for oxygen emissions, this response was not restricted to the peak layer but extended over a wide altitude range (70–200 km). Furthermore, the relative enhancement of the 135.6 nm emission compared to the 130.4 nm emission was generally stronger, particularly at lower SZAs (Figure 2f).

To complement the IUVS limb-observations and investigate the large-scale horizontal extent of the ICME response, we utilize EMUS disk-observations (Figure 3). It represents EMUS disk images of H Lyman-α, O($^3$S) 130.4 nm, and O($^5$S) 135.6 nm emissions during quiet-time (left column) and ICME conditions (middle column), together with spatial profiles (right column) derived from the blue and red boxed regions. Across all panels, the *x*-axis represents the readout index (integration period), while the *y*-axis shows spatial bins in the left and middle columns, and radiance (Figures 3c, 3f, 3i) or SZA (Figure 3l) in the right column. The EMUS observation during this ICME event (18 May 2024 at 09:59 UT) closely coincided with IUVS observations from orbit 21225 (18 May 2024 at 09:29 UT), enabling near-simultaneous characterization of the ICME impact. For quiet-time reference, we selected the EMUS disk scan acquired on 20 April 2024 at 14:47 UT to ensure a comparable viewing geometry and similar dayside SZA coverage (Figure S2 in Supporting Information) to that of the event time. During both quiet and event periods, EMUS sampled the southern hemispheric dayside disk region over nearly identical SZA ranges (Figure 3l), thereby minimizing geometrical biases in the comparison. A consistent enhancement was observed by EMUS (Figure 3) in all three emissions across the dayside disk during the ICME. Relative to quiet time, event-time H Lyman-α near the subsolar region increased from ~15–18 R to ~25–30 R (Figure 3c), O($^3$S) 130.4 nm from ~1000–1100 R to ~1400–1500 R (Figure 3f), and O($^5$S) 135.6 nm increased from ~250–300 R to ~350–400 R (Figure 3i). The SZA maps (Figures 3j–3l) further show that the enhancement extended across a broad range of dayside SZAs and toward the terminator. Together with the IUVS limb profiles, these observations demonstrate that the ICME response is both horizontally extensive and vertically distributed.

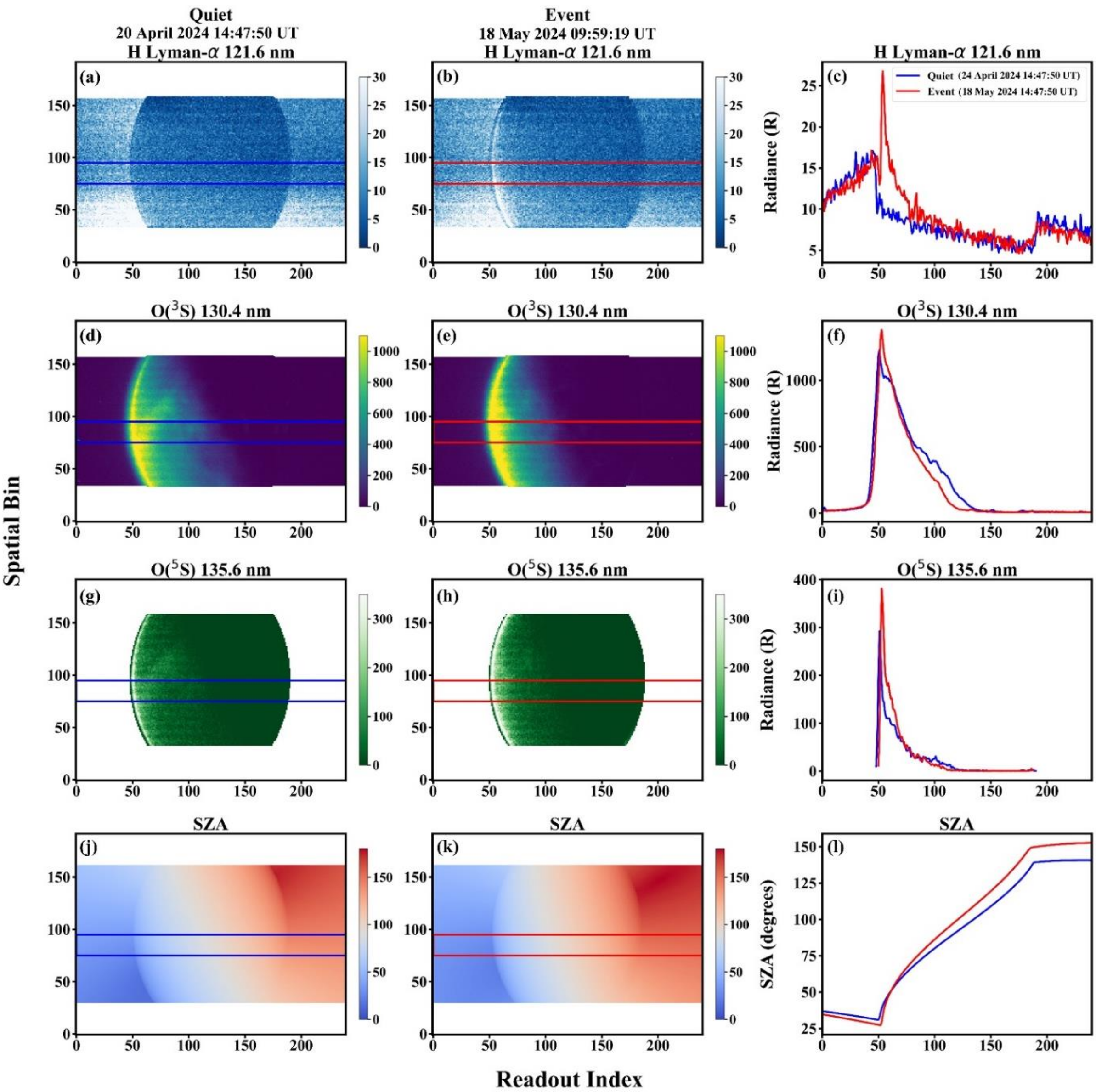

**Figure 3:** EMUS disk images of H Lyman-α, O(³S) 130.4 nm, and O(⁵S) 135.6 nm under quiet (left) and ICME (middle) conditions, with spatial profiles (right) extracted from blue (quiet) and red (event) boxes. Bottom panels show SZA variation.

Figure 4 presents EMUS dayglow brightness variations across latitude (left), longitude (middle), and local-time (right) for different SZA bins. Data were restricted to the peak emission layer (130–150 km). Solid and dashed lines denote event-time and quiet-time conditions, respectively, with shaded regions indicating the variability within each bin. Across all emissions, the ICME produced systematic and spatially coherent enhancements relative to quiet-time. The H Lyman-α emission (Figure 4a) shows the strongest response, with event-time brightness exceeding quiet-time by ~80–200% in the low SZA bins (20°–40°), particularly at mid-latitudes (−50° to −30°N) where intensities increase from ~15–20 R to ~40–50 R. This enhancement has decreased with increasing SZA but remained elevated by ~20–50% even in higher SZA bins (60°–70°). The longitudinal distribution (Figure 4b) exhibits pronounced peak near 50°–100°E. Whereas, the local-time distribution (Figure 4c) shows maximum enhancement between 10-12 h, corresponding to the subsolar region, with enhancement extending into afternoon sector. The $O(^3S)$ 130.4 nm emission (Figures 4d–4f) exhibits a similar enhancement, with event-time brightness increasing by ~20–50% relative to quiet-time across most spatial bins. Peak intensities increased from ~900–1100 R (quiet) to ~1200–1400 R (event) at mid-latitudes and near subsolar longitudes. The spatial distribution closely follows solar illumination, with maximum enhancement near ~100°–150°E longitude and local-time ~10–12 h, and a gradual decrease toward higher SZA and evening sectors. The $O(^5S)$ 135.6 nm emission (Figures 4g–4i) shows a stronger enhancement (~40–80%), especially at lower SZA, compared to quiet-time values. The peak brightness increased from ~300–400 R to ~400–500 R at mid-latitudes. The spatial patterns in longitude and local time closely resemble those of the 130.4 nm emission but exhibit consistently larger relative enhancements. Consistent across all three emissions, the ICME-driven enhancement was greatest at low SZAs and gradually weakened toward the terminator, although brightness remained above quiet-time levels throughout the observed dayside. This behavior is consistent with the IUVS limb observations (Figure 2), which show enhancement across a broad SZA range.

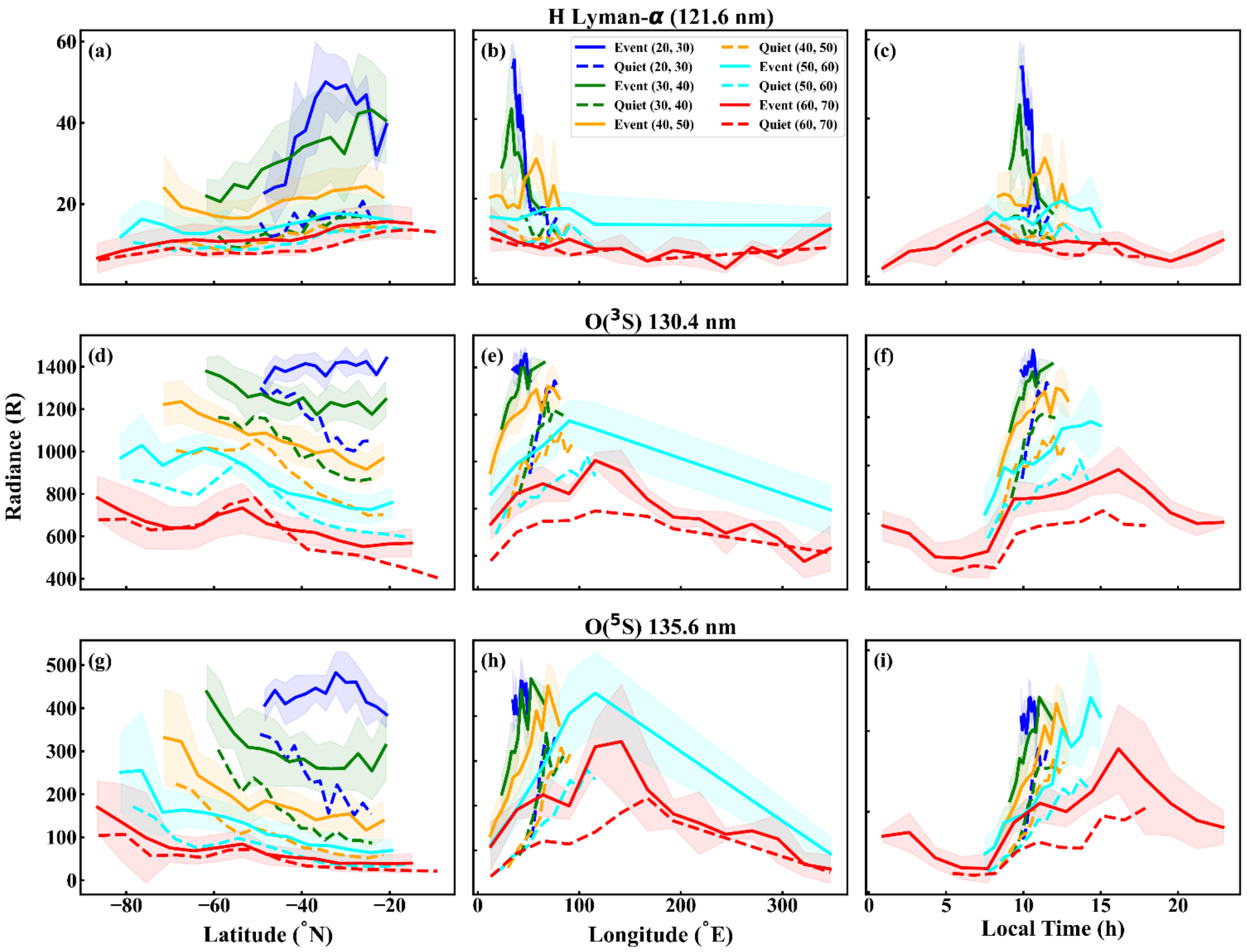

**Figure 4:** Variation of H Lyman-α, $O(^3S)$ 130.4 nm, and $O(^5S)$ 135.6 nm brightness with latitude (left), longitude (middle), and local-time (right) for different SZA bins (indicated in the legend). All data are taken from the 130–150 km altitude range. Solid curves represent event-time conditions, dashed curves quiet-time conditions, and shaded regions denote 1σ variability.

**4. Discussion and Conclusion**

This study investigates the impact of May 2024 ICME event on Martian H Lyman-α, O($^3$S) 130.4 nm and O($^5$S) 135.6 nm dayglow emissions. Past studies have characterized ICME-driven dayglow enhancements using localized limb observations (Deighan et al., 2018; Ritter et al., 2018; Sharma et al., 2025), limiting our understanding of the large-scale atmospheric response. We utilized near-simultaneous MAVEN/IUVS limb and EMM/EMUS disk observations providing a unique opportunity to examine the ICME response in both the vertically and horizontally in the Martian upper atmosphere. In Figure 2, IUVS observations show significant enhancements in H Lyman-α, O($^3$S) 130.4 nm, and O($^5$S) 135.6 nm emissions during the ICME, with peak brightnesses that reached ~32 kR, ~2.46 kR, and ~2.56 kR, respectively. Complementary EMUS observations (Figures 3&4) independently confirm a broad large-scale enhancement, where H Lyman-α reached ~25–30 R, O($^3$S) 130.4 nm ~1400–1500 R, and O($^5$S) 135.6 nm ~350–400 R during the event. These observations coincided with substantial increases in solar wind parameters, **|B|** strength, and SEP fluxes (Figure 1), indicating strong coupling between the ICME and the Martian thermosphere. The following discussion examines the physical mechanisms responsible for these observed enhancements.

The H Lyman-α enhancement observed during the May 2024 ICME is consistent with the proton aurora mechanism proposed by Deighan et al. (2018), in which solar wind protons undergo charge-exchange with exospheric hydrogen to form ENAs that subsequently precipitate into the atmosphere and excite hydrogen emissions (R6). Previous studies have shown that proton auroral brightness increases during periods of enhanced solar wind velocity, particle flux, and dynamic pressure (Hughes et al., 2023, 2025, 2026), while Hughes et al. (2024) further demonstrated a positive correlation between proton auroral intensity and IMF magnitude (|**B**|). Consistent with these findings, the May 2024 ICME was characterized by a substantial increase in solar wind density, velocity, dynamic pressure, and |**B**| strength (Figure 1), coinciding with a pronounced enhancement in H Lyman-α brightness observed by both IUVS and EMUS. In addition, the elevated H Lyman-α intensity persisted across multiple consecutive IUVS orbits (Figure 2a–e), likely due to multi-day sustained high solar wind velocity and density that drove increased precipitation of $H^+$ and H at thermospheric altitudes (Hughes et al., 2023). Moreover, the present observations were acquired during Martian southern summer (Ls ~270°), when the expanded hydrogen corona favours the most frequent and intense proton aurora events (Hughes et al., 2019). Together, the favourable seasonal conditions and extreme ICME forcing likely contributed to the exceptionally strong H Lyman-α enhancement observed during this event.

While the H Lyman-α enhancement reflects proton-driven processes, the oxygen emissions provide insight into electron-driven excitation mechanisms. Interestingly during this event, both IUVS and EMUS also observed enhancement in $O(^3S)$ 130.4 nm and $O(^5S)$ 135.6 nm dayglow emissions. The simultaneous but spatially localized enhancement of H Lyman-α, 130.4 nm, and 135.6 nm dayglow emissions was previously reported by Chaffin et al. (2022), where coincident H and O dayglow emission enhancements were attributed to energy deposition by precipitating $H^+$ beams along the open magnetic field lines. In contrast, present event exhibits substantial oxygen dayglow enhancement over a broad range of altitudes, latitudes, longitudes, local-times, and SZAs. The increased oxygen emissions were likely driven by enhanced solar wind electron flux (Figure 1e), SEP electron and ion energy fluxes (Figure 1f&1g), which increased to $\sim 2\times10^8$, $\sim 1.5\times10^4$, and $\sim 5\times10^4$ eV $cm^{-2}$ $s^{-1}$ $sr^{-1}$, respectively. This large increase in the electron/ion fluxes enhances ionisation that can further produce secondary electrons at thermospheric altitudes (Nakamura et al., 2022; Felici et al., 2024). In addition, energetic solar wind and SEP electrons can directly impact $CO_2$ molecules. Together, these processes can augment the electron-impact processes (R2-R5; Sharma et al., 2025), leading to increased dayglow brightness.

The comparable or slightly stronger enhancement of $O(^5S)$ 135.6 nm (~2.52 kR) relative to $O(^3S)$ 130.4 nm (~2.46 kR) suggests an increased contribution from electron-impact excitation (Figure 2). Further insight is provided by the normalized oxygen emission ratio (Figure S6 in Supporting Information). Ritter et al. (2019) showed that $O(^3S)$ 130.4 nm emission is dominated by resonance scattering, whereas $O(^5S)$ 135.6 nm emission is primarily produced through electron-impact excitation. Consequently, the normalized [$O(^3S)$130.4/$O(^5S)$135.6] ratio serves as a diagnostic of photon-driven (R1) versus particle-driven processes (R2-R5). The ratio remains consistently below unity across all scans and spatial bins, reaching values as low as ~0.5 during the peak ICME phase. Importantly, the solar EUV flux did not increase during the event compared to quiet-time (Figure S7 in Supporting Information). This indicate that solar irradiance is unlikely to explain the stronger relative enhancement of the electron-impact-sensitive $O(^5S)$ 135.6 nm emission. Instead, the reduced ratio indicates enhanced electron precipitation, consistent with the elevated solar wind and SEP electron fluxes observed during the event (Figure 1e–g). Soret et al. (2024) further showed that lower [$O(^3S)$130.4/$O(^5S)$135.6] ratios correspond to harder precipitating electron spectra. The observed ratios (~0.5–0.7) therefore suggest both enhanced and relatively energetic electron precipitation during the ICME. Orbit 21225 exhibits the lowest ratio and coincided with the EMUS observation, confirming that both instruments captured the strongest phase of the electron-driven response.

Beyond brightness increases, the near-simultaneous IUVS (orbit 21225) and EMUS disk observations demonstrate that the oxygen enhancement also extended in vertical and spatial dimensions. Under quiet conditions, the oxygen emission peaks are largely confined between ~120–140 km. However, during ICME, the emission peak broadened to ~100–160 km (Figure 2) while maintaining a peak altitude similar to quiet-time. Concurrently, EMUS shows enhanced oxygen emissions across a broad spatial range (Figures 4). Furthermore, the increase was also observed both above and below the emission peak, indicating that the ICME affected the entire thermospheric column rather than a narrow altitude region. This is supported by the increased SEP energy flux (Figure 1f) and solar-wind electron flux (Figure 1e) observed during ICME. The altitude structure provides important insight into the precipitating particle energy distribution. Previous electron transport studies had shown that lower-energy electrons deposit energy at higher altitudes, whereas higher energy electrons penetrate deeper into the atmosphere (Lillis and Fang, 2015; Jolitz et al., 2021). They showed that electrons with energies of a few hundred eV primarily deposit energy near ~140–170 km, while keV to 10s of keV electrons can penetrate to ~80–120 km. The observed broadening therefore suggests simultaneous precipitation of a wide spectrum of particle energies.

The vertical broadening observed in IUVS limb profiles is complemented by EMUS disk observations, which demonstrate that the ICME response extends across much of the observed dayside thermosphere. Xu et al. (2018) showed that strong ICME conditions can allow IMF to penetrate below ~300 km altitude, where open magnetic field topologies are formed through the magnetic-reconnection. This facilitates energetic particles access to the thermosphere over broad spatial regions. Consistent with this scenario, the hydrogen and oxygen emissions show the strongest enhancement at low SZA (<50°) and gradually decrease toward the terminator (Figure 4). This reflects stronger photoionization and photoelectron production near the subsolar region. However, the IUVS limb profiles continue to show substantial enhancement at higher SZAs, whereas this behaviour is less evident in the EMUS observations (Figure 2&4). This enhancement observed over terminator can be linked to solar wind conditions, which can enhance suprathermal electron precipitation and day-to-night plasma transport near the terminator and nightside regions (Krishnaprasad et al., 2019; Xu et al., 2020; Kajdič et al., 2021), potentially contributing to the persistent high-SZA enhancement observed by IUVS.

Furthermore, we find that the latitude, longitude, and local-time distributions (Figure 4) follow the same trend as observed during quiet-time conditions, with the ICME primarily increased brightness rather than reorganizing the spatial pattern of dayglow emissions. Similarly, the IUVS limb profiles (Figure 2) show that the nominal peak altitudes remain largely unchanged despite substantial increases in intensity. Together, these observations indicate that the ICME enhanced the brightness of the dayglow emissions while largely preserving their characteristic spatial, local-time, and SZA distributions. Similar enhancement without significant changes in peak altitude have been reported during solar-flare-driven dayglow events (Ram et al., 2024) and for the Martian $O(^1S)$ 557.7 nm dayglow emission during ICME (Sharma et al., 2025). While those studies were limited to limb geometry, the combined IUVS and EMUS observations presented here demonstrate that this pattern remains unchanged and occurred simultaneously across both altitude and broad spatial scales.

In conclusion, the combined MAVEN/IUVS and EMM/EMUS observations demonstrate that the May 2024 ICME produced a strong, vertically distributed and spatially extensive enhancement in Martian UV dayglow emissions. The near-simultaneous IUVS limb and EMUS disk observations provide the first of its kind investigation of the ICME response in both the vertical and horizontal dimensions of the Martian upper atmosphere. The oxygen emissions exhibited substantial vertical broadening, indicating distributed energy deposition across the thermosphere by a broad spectrum of precipitating electrons, while the persistently reduced [$O(^3S)$130.4/$O(^5S)$135.6] ratio suggests an enhanced contribution from electron-impact excitation. The distinct spatial responses of H Lyman-α and oxygen emissions demonstrate that the ICME simultaneously enhanced proton-driven and electron-driven excitation pathways. Furthermore, the latitude, longitude, local-time distributions, and emission peak altitudes remain largely unchanged, indicating that the ICME primarily enhanced brightness while preserving the characteristic altitude and spatial structure of the UV dayglow emissions.

**Acknowledgments**
We sincerely acknowledge the NASA PDS, EMM SDC, MAVEN and EMM team members for the datasets. A. R. Shamra acknowledges the fellowship from the Ministry of Education, Government of India. L. Ram acknowledges the fellowship from the Japan Society for Promotion of Science (JSPS) for supporting this research. This work is supported by the Ministry of Education, Government of India.

**Data Availability Statement**

The NASA DONKI catalogue was utilized for space weather forecasting. The corresponding ICME details are available at https://kauai.ccmc.gsfc.nasa.gov/DONKI/view/CMEAnalysis/30868/3, and the simulation results, including relevant images and GIFs, can be accessed at https://kauai.ccmc.gsfc.nasa.gov/DONKI/view/WSA-ENLIL/30875/1. MAVEN data set of SWIA (solar-wind ion) Level 2, version_19, revision_01, MAG (IMF) Level 2, version_19, revision_01, SWEA (electron-flux) Level 2, version_19, revision_01, and EUV Monitor Level 3, version_15, revision_01 used in this work are obtained from the NASA Planetary Data System (PDS) and can be accessed through https://pds-ppi.igpp.ucla.edu/search/default.jsp. The EMM/EMUS l2b datasets used in this work are available at the EMM Science Data Center (SDC, http://sdc.emiratesmarsmission.ae). The derived data products used in this study can be found in Sharma et al. (2026).